\documentclass[preprint,amsmath,amssymb,aps,nofootinbib,showpacs]{revtex4}
\usepackage[utf8]{inputenc}
\usepackage{graphicx}
\usepackage{bm}
\usepackage{subfigure}
\usepackage{amsmath}

\newcommand{\pr}{Phys. Rev.\ }

\newcommand{\jpb}{J. Phys. B\ }
\newcommand{\njp}{New J. Phys.\ }
\newcommand{\etal}{{\em et al. }}
\newcommand{\e}{\mbox{e}}

\newcommand{\UQ}{School of Mathematics and Physics, University of Queensland, Brisbane, 
QLD 4072, Australia.}

\begin{document}

\title{Pseudo steady-state non-Gaussian EPR-steering of massive particles in pumped and damped Bose-Hubbard dimers}

\author{M.~K. Olsen}
\affiliation{\UQ}

\date{\today}

\begin{abstract}

We propose and analyse a pumped Bose-Hubbard dimer as a source of continuous-variable Einstein-Podolsky-Rosen (EPR) steering with non-Gaussian statistics. We use the truncated Wigner representation to calculate third and fourth order cumulants, finding clear signals of non-Gaussianity. We also calculate the products of inferred quadrature variances which indicate that states demonstrating the EPR paradox are present. Our proposed experimental configuration is extrapolated from current experimental techniques and adds another possibility to the current toolbox of quantum atom optics.

\end{abstract}

\pacs{03.75.Lm,03.75.Gg,03.65.Yz}       

\maketitle


Continuous-variable (CV) systems provide flexible and
powerful means for implementing quantum-information
schemes~\cite{Braunstein}, in large part this is because there are mature and
precise techniques for measuring the quadratures of light.
Most of these are familiar from classical communications
technologies, and are being extended to atomic measurements~\cite{Andy}.
One stumbling block for the wider use of optical CV
systems is that the most-readily available CV systems and the
most-developed detection techniques produce only Gaussian
statistics. This limitation rules out tasks such as entanglement
distillation~\cite{Eisert} and quantum error correction~\cite{Niset}. One way of introducing non-Gaussian statistics
is through nonlinear measurements~\cite{Lvovsky}, but this approach
negates the advantages of the
highly developed technology that is available for performing
Gaussian homodyne measurements. It is therefore of interest to analyse the production of non-Gaussian entanglement and Einstein-Podolsky-Rosen (EPR) states~\cite{EPR} directly through non-linear interactions such as found in optical fibres~\cite{nonGauss}, so that homodyne detection technology may be used. 

It is well known that such nonlinear interactions are also present in condensed bosons, where non-Gaussian states arise naturally from $\chi^{(3)}$ nonlinear processes via the s-wave collisional interaction.
The manufacture of non-Gaussian entangled states has been analysed with massive particles, both theoretically~\cite{KVKEPR,MDRbiwell,BobBell}, and experimentally~\cite{Oberthaler,EPRatoms}. 
Along with the available measurement techniques, recent advances in the technology of optical potentials~\cite{painting,tylerpaint} allow for an increased flexibility in the trapping and measurement of ultra-cold gases. Combined with dissipation from a particular well via the use of either an electron beam~\cite{NDC}, or by optical means~\cite{Weitenberg}, and the possibility of pumping a Bose-Hubbard system from a larger reservoir condensate~\cite{Kordas1,Kordas2}, we have new opportunities for the fabrication of nonlinear damped and pumped atom-optical equivalents of optical cavities with varying configurations~\cite{BHcav2,Stuttgart}. 
In this work we investigate two different Bose-Hubbard models~\cite{BHmodel,Jaksch,BHJoel} with added pumping and loss, in terms of their utility for the preparation of non-Gaussian EPR states of the two atomic modes, quantifying both the non-Gaussianity of the resulting quantum states and the degree of violation of standard EPR inequalities~\cite{EPRMDR}.


Our systems consist of two wells, each able to contain a single atomic mode. They both have pumping into the first well and differ in which well is damped. 
To describe them, we begin with the two-well unitary Bose-Hubbard Hamiltonian~\cite{BHJoel}, written as
\begin{equation}
{\cal H} = \hbar\chi\sum_{i=1}^{2}\hat{a}_{i}^{\dag\,2}\hat{a}_{i}^{2}-\hbar J \left(\hat{a}_{1}^{\dag}\hat{a}_{2}+\hat{a}_{2}^{\dag}\hat{a}_{1} \right),
\label{eq:genHam}
\end{equation}
where $\hat{a}_{i}$ is the bosonic annihilation operator for the $i$th well, $\chi$ represents the collisional nonlinearity and $J$ is the tunneling strength. We will consider that the pumping into well $1$ and can be represented by the Hamiltonian
\begin{equation}
{\cal H}_{pump} = i\hbar\left(\hat{\Gamma}\hat{a}_{1}^{\dag}-\hat{\Gamma}^{\dag}\hat{a}_{1}\right),
\label{eq:pump}
\end{equation}
which is the same form as that commonly used for the investigation of pumped optical cavities. The basic assumption here is that the first well receives atoms from a coherent condensate which is much larger than any of the modes in the wells we are investigating, so that it will not become noticeably depleted over the time scales of interest. The pumping condensate is the equivalent of the pumping laser used with optical cavities.
The damping term for well $i$ acts on the system density matrix as the Lindblad superoperator
\begin{equation}
{\cal L}\rho = \gamma\left(2\hat{a}_{i}\rho\hat{a}_{i}^{\dag}-\hat{a}_{i}^{\dag}\hat{a}_{i}\rho-\rho\hat{a}_{i}^{\dag}\hat{a}_{i}\right),
\label{eq:damp}
\end{equation}
where $\gamma$ is the coupling between the damped well and the atomic bath, which we assume to be unpopulated. Physically, such a damping process can be realised using either an electron beam~\cite{NDC} or by optical methods~\cite{Weitenberg}. With the outcoupled atoms falling under gravity, we are justified in using the Markov and Born approximations~\cite{JHMarkov}.

With the above Hamiltonians and the Lindblad superoperator as our starting point, there are several possible ways in which we could proceed.We could choose density matrix techniques, as used by Pi\u{z}orn~\cite{Pizorn}, which are useful for moderate numbers of atoms and wells. However, we do not wish to work in a truncated Hilbert space and have no a priori knowledge of what either the final or transient quantum states may be, so that we would not know in advance how to safely carry out any truncation. Cui \etal have investigated driven and dissipative Bose-Hubbard models, obtaining mean-field analytical results for a two-well system~\cite{Cui}, but we wish to calculate quantum correlations, which is not possible with a mean-field analysis. As in previous work~\cite{BHcav2,Chiancathermal,CTAP}, we will use the truncated Wigner representation~\cite{Graham,Steel}, which does not impose a computational limitation on the number of atoms. Although this method is also an approximation, effectively including the first quantum corrections of a $(1/N)$ expansion, we fully expect it to be accurate for our systems of mixed states in the presence of pumping and dissipation. The truncated Wigner representation goes beyond the pairing mean-field theory~\cite{PMFT} and the Bogoliubov back reaction method~\cite{BBRVardi1,BBRVardi2} previously used in theoretical analyses in that it imposes no factorisation assumptions on correlations, irrespective of their order. It has also been shown that, for a single mode system, the truncated Wigner representation is able to reproduce the non-Gaussian correlations~\cite{WigJFC} and that it will closely reproduce the occupation probabilities in a two-well Bose-Hubbard model~\cite{WigBob}.

Following the usual procedures~\cite{QNoise,DFW}, we map the problem onto a generalised Fokker-Planck equation (FPE) for the Wigner distribution of the system. Since this generalised FPE contains third-order derivatives and these cannot be mapped onto stochastic differential equations, we truncate at second order. Having discarded these derivatives, we may map the resulting FPE onto It\^o stochastic equations~\cite{SMCrispin} for the Wigner variables. These equations for a two-well chain with pumping at well $1$ and loss at well $2$ are
\begin{eqnarray}
\frac{d\alpha_{1}}{dt} &=& \epsilon - 2i\chi |\alpha_{1}|^{2}\alpha_{1}+iJ\alpha_{2}. \nonumber \\
\frac{d\alpha_{2}}{dt} &=& -\gamma\alpha_{2}-2i\chi|\alpha_{2}|^{2}\alpha_{2}+iJ\alpha_{1}+\sqrt{\gamma}\eta,
\label{eq:BHp1g2}
\end{eqnarray}
with those with loss at the pumped well resulting from moving the terms involving $\gamma$ into the first equation and changing the appropriate index. In the above equation, $\epsilon$ represents the rate at which atoms enter well $1$ from the pumping mode, $\gamma$ is the atomic loss rate, and $\eta$ is a complex Gaussian noise with the moments $\overline{\eta(t)}=0$ and $\overline{\eta^{\ast}(t)\eta(t')}=\delta(t-t')$, where the upper line represents a classical averaging process. The variables $\alpha_{i}$ correspond to the operators $\hat{a}_{i}$ in the sense that averages of products of the Wigner variables over many stochastic trajectories become equivalent to symmetrically ordered operator expectation values, for example $\overline{|\alpha_{i}|^{2}}=\frac{1}{2}\langle\hat{a}_{i}^{\dag}\hat{a}_{i}+\hat{a}_{i}\hat{a}_{i}^{\dag}\rangle$. The initial states in both wells are vacuum. We note here that we will use $\epsilon=10$ and $\gamma=J=1$ in all our numerical investigations, while varying the value of $\chi$.
  

In order to investigate the non-Gaussian nature of the system, we first define atomic quadratures as 
\begin{eqnarray}
\hat{X}_{j}(\theta) &= & \hat{a}_{j}\e^{-i\theta}+\hat{a}_{j}^{\dag}\e^{i\theta}.
\label{eqn:Xtheta}
\end{eqnarray}
This allows us to calculate the third and fourth-order quadrature cumulants,  $\kappa_{3}$ and $\kappa_{4}$~\cite{nonGauss,SMCrispin}, defined for either of the $\hat{X}_{j}$ quadratures as,
\begin{eqnarray}
\kappa_{3}(\hat{X}) &=& \langle \hat{X}^{3} \rangle + 2\langle\hat{X} \rangle^{3} - 3\langle\hat{X} \rangle\langle\hat{X}^{2}\rangle, \nonumber \\
\kappa_{4}(\hat{X}) &=&  \langle \hat{X}^{4} \rangle + 2\langle\hat{X} \rangle^{4} - 3\langle\hat{X}^{2}\rangle^{2}-\langle\hat{X}\rangle\kappa_{3}(\hat{X}),
\label{eq:kapp3&4}
\end{eqnarray}
where we have suppressed the quadrature angles for clarity of notation.
A non-zero value of either of these is sufficient to demonstrate that the statistics of the system are non-Gaussian.
 
The presence of EPR-steering~\cite{EPR,Erwin,Jonesteer} is signified by violation of the Reid inequalities for the inferred variances~\cite{EPRMDR}
\begin{equation}
\Pi V_{ij} = V^{inf}(\hat{X}_{i})V^{inf}(\hat{Y}_{i})\geq 1,
\label{eq:eprMDR}
\end{equation}
where
\begin{eqnarray}
V_{inf}(\hat{X}_{i}) &=& V(\hat{X}_{i})-\frac{[V(\hat{X}_{i},\hat{X}_{j})]^{2}}{V(\hat{X}_{j})}, \nonumber \\
V_{inf}(\hat{Y}_{i}) &=& V(\hat{Y}_{i})-\frac{[V(\hat{Y}_{i},\hat{Y}_{j})]^{2}}{V(\hat{Y}_{j})},
\label{eq:EPRdef}
\end{eqnarray}
and $V(\hat{A},\hat{B})=\langle\hat{A}\hat{B}\rangle-\langle\hat{A}\rangle\langle\hat{B}\rangle$.
The presence of an EPR state is indicated by violation of the inequality.
This condition is optimal for bipartite Gaussian systems, and at least sufficient for non-Gaussian systems.


For our two-well damped and driven dimer, we investigate two different configurations. The first has pumping at well $1$ with loss at well $2$, while the second has both pumping and loss at well $1$. As we have shown previously~\cite{BHcav2}, these configurations exhibit qualitatively different behaviours in terms of the population dynamics, with both having mesoscopic steady-state occupations of the two wells. As we show below, both are possible sources of non-Gaussian CV states of massive particles which exhibit EPR-steering. 

\begin{figure}[tbhp]
\includegraphics[width=0.75\columnwidth]{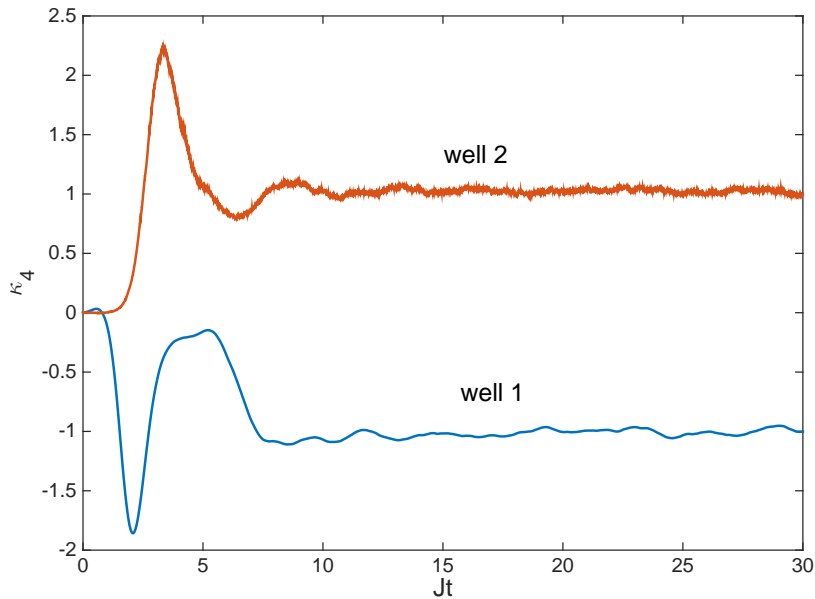}
\caption{(colour online) Averaged values for $\kappa_{4}$ in the two wells, with $\chi=10^{-3}$, $\epsilon=10$, $\gamma=1$, and $J=1$. Damping is at the second well and the horizontal axis is a dimensionless time, $Jt$. All quantities plotted here and in subsequent figures are dimensionless.}
\label{fig:kap4g2ki3}
\end{figure}

\begin{figure}[tbhp]
\includegraphics[width=0.75\columnwidth]{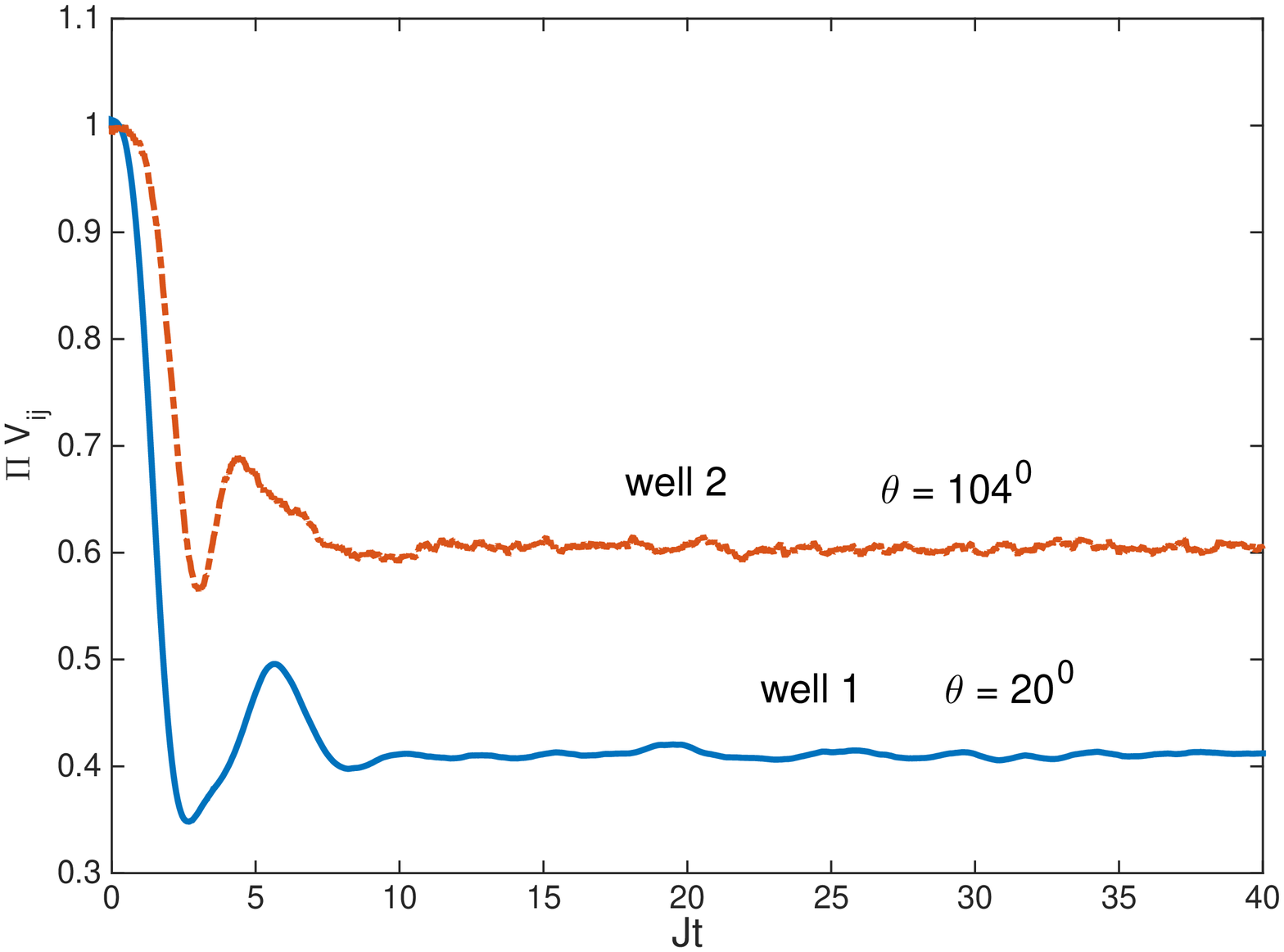}
\caption{(colour online) The products of the inferred quadrature variances for damping at well $2$, with $\chi=10^{-3}$, $\gamma=1$, $\epsilon=10$ and $J=1$, optimised for the quadrature angles of greatest violation of the inequalities. The value marked well $i$ refers to the product $\Pi V_{ij}$.}
\label{fig:EPRg2ki3}
\end{figure}

The first configuration is described by Eq.~\ref{eq:BHp1g2}. We found that the results for the cumulants typically needed averaging over a far greater number of trajectories to converge than was required for either the populations or the EPR correlations. While this is not unexpected since they contain higher order operator moments, the results of Fig.~\ref{fig:kap4g2ki3} required $12.6\times 10^{6}$ trajectories, as against something of the order of $10^{5}$ for lower order correlations. This figure 
shows the results for the fourth-order cumulant, $\kappa_{4}$ at $\theta=0$, for a nonlinear interaction value of $\chi=10^{-3}$, demonstrating clearly that the statistics of the system are non-Gaussian. We note here that either of $\kappa_{3}\neq 0$ or $\kappa_{4}\neq 0$ is sufficient to reveal non-Gaussian statistics, but in this case $\kappa_{4}$ gives a much more definite signal. We also note that, by this measure, increasing the nonlinearity increased the degree of non-Gaussianity. We did not optimise the quadrature angles to find the largest magnitude value of the cumulants, because there is a clear signal at $\theta=0$.

\begin{figure}[tbhp]
\includegraphics[width=0.75\columnwidth]{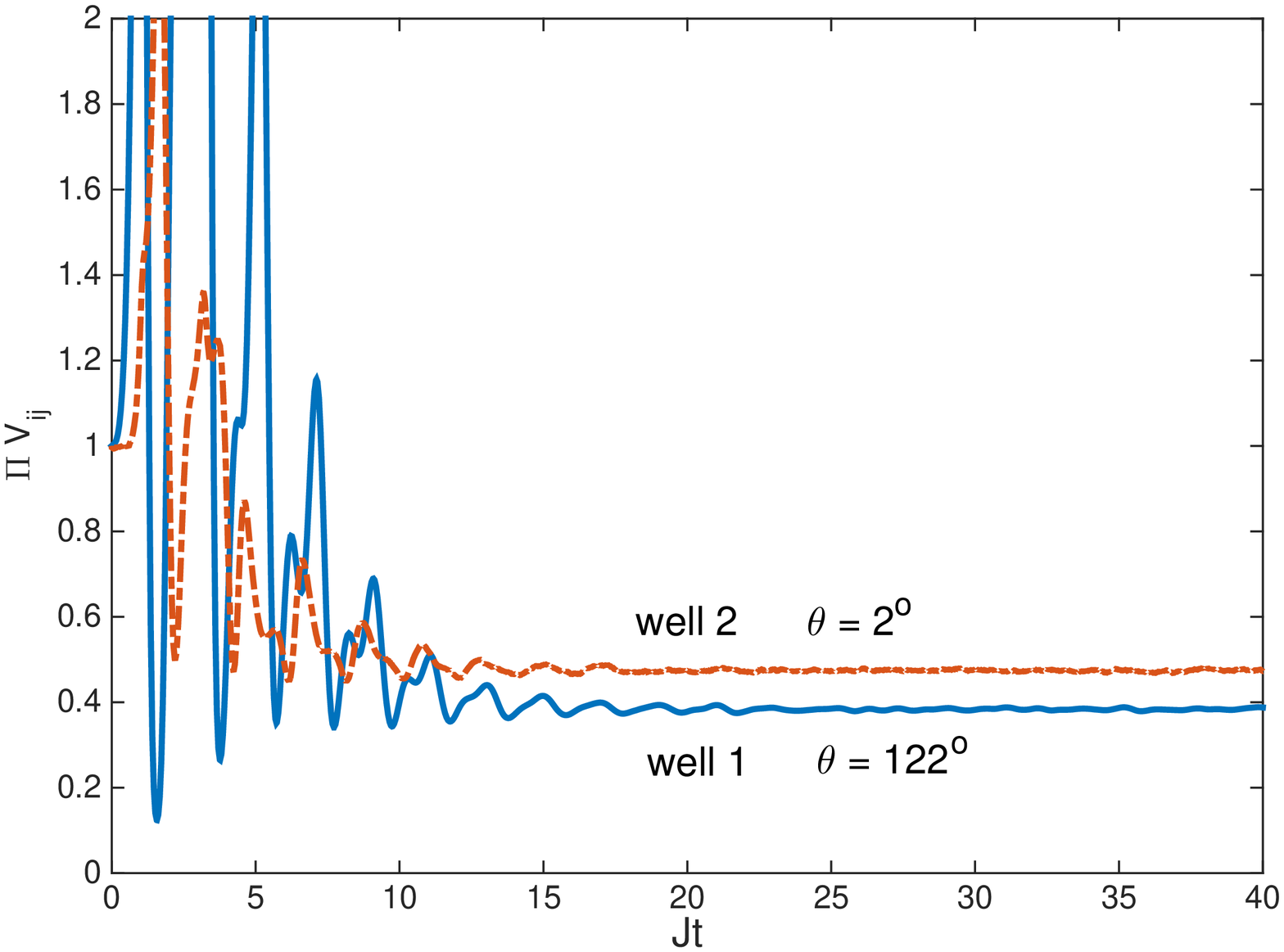}
\caption{(colour online) The products of the inferred quadrature variances for damping at well $2$ for $\chi=10^{-2}$, $\gamma=1$, $\epsilon=10$, and $J=1$, optimised for the quadrature angles of greatest violation of the inequalities. The value marked well $i$ refers to the product $\Pi V_{ij}$.}
\label{fig:EPRg2ki2}
\end{figure}

In Fig.~\ref{fig:EPRg2ki3} we show clear violations of the EPR-steering inequalities for $\chi=10^{-3}$, with the two different results depending on which well is used to infer the variances of the other. A value for these of less than $1$ demonstrates that continuous-variable EPR-steering is present in the system. In Fig.~\ref{fig:EPRg2ki2} we show that the steady-state EPR-steering violations increase when $\chi$ is increased to $10^{-2}$. At the same time the statistics become more non-Gaussian, but increasing the collisional nonlinearity further would take us beyond the limits of our model. We see that the maximal violations are asymmetric as regards the optimal quadrature angles, although to demonstrate true asymmetric steering~\cite{Sarah}, there would need to be no violation in one of the wells for any quadrature angle, while a violation at some quadrature angle existed in the other.
It is interesting to note here that we only found marginal violations of the Duan-Simon inequalities for these systems~\cite{Duan,Simon,BHcav2}. This can be explained by the fact that the Duan-Simon correlations work optimally to detect entanglement and inseparability in pure Gaussian systems, while the Reid EPR correlations do not have this limitation and work well in our mixed state non-Gaussian systems.

\begin{figure}[tbhp]
\includegraphics[width=0.75\columnwidth]{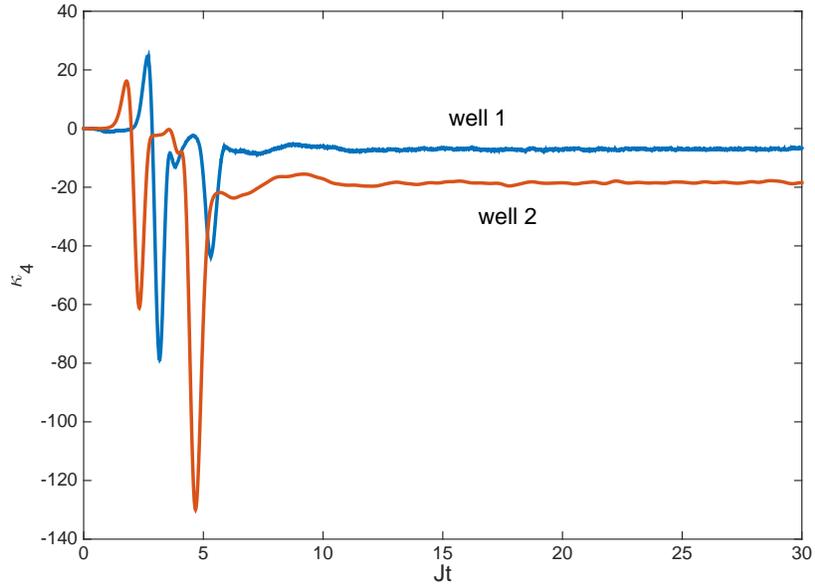}
\caption{(colour online) Averaged values for $\kappa_{4}$ in the two wells, with $\chi=10^{-2}$, $\epsilon=10$, $\gamma=1$, $J=1$, and damping at the first well.}
\label{fig:kap4g1ki2}
\end{figure}

The second configuration has both pumping and dissipation at the first well and can be realised by moving either the electron beam or the damping laser onto the pumped well. As shown in Fig.~\ref{fig:kap4g1ki2}, this system is also a source of non-Gaussian states in the steady-state. The values for $\chi=10^{-3}$ also give a clear non-Gaussian signal. $\kappa_{3}$ was also non-zero in both cases, but does not give as large a signal. SInce only one cumulant of higher than second order need be non-zero to denote non-Gaussian states, we have chosen to show only the strongest of the two. Interestingly, these results converged well with averaging over $5\times 10^{5}$ trajectories, an order of magnitude less than with damping at the second well.

\begin{figure}[tbhp]
\includegraphics[width=0.75\columnwidth]{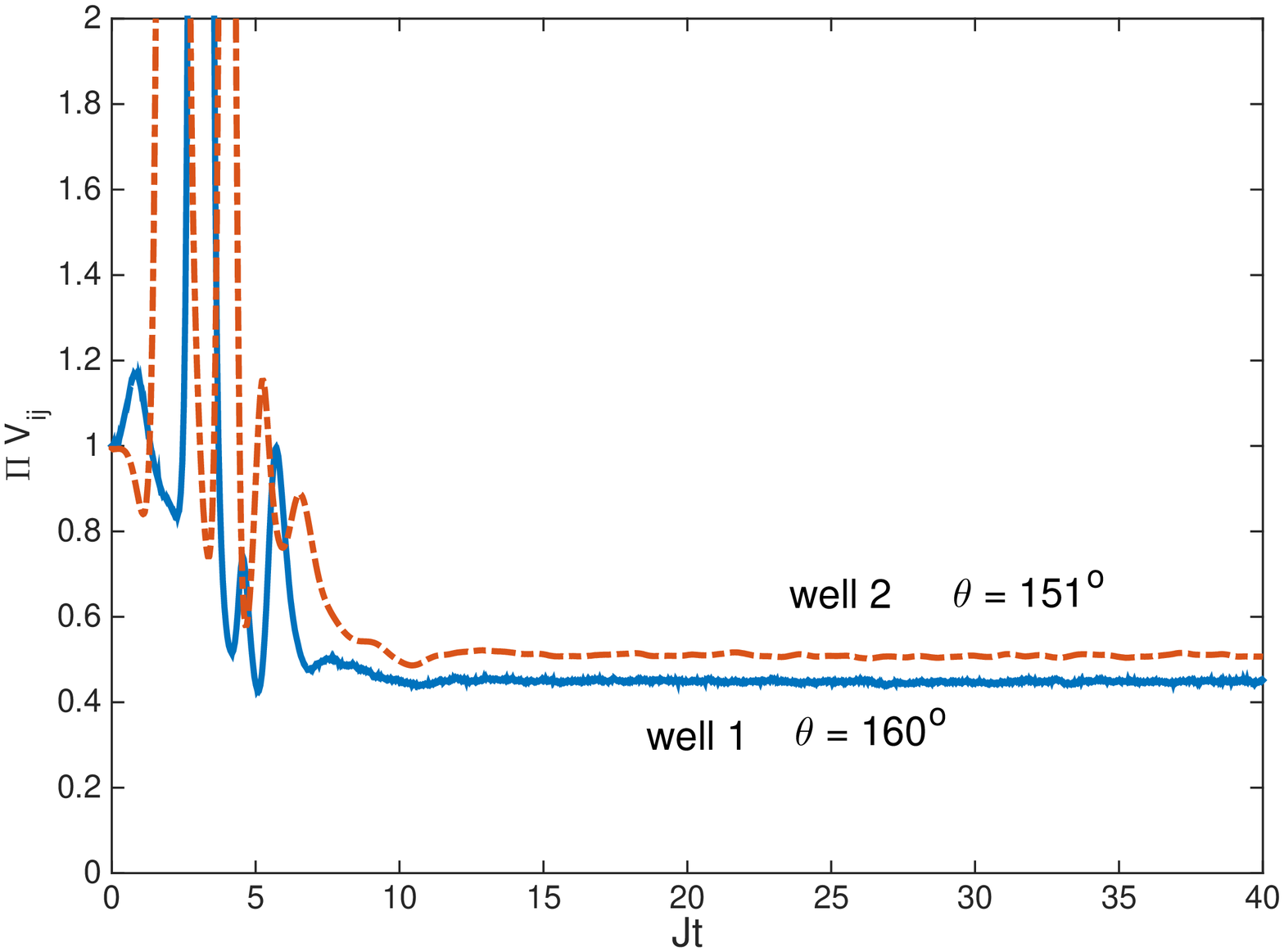}
\caption{(colour online) The products of the inferred quadrature variances for damping at well $1$, with $\chi=10^{-2}$, $\gamma=1$, $\epsilon=10$ and $J=1$, optimised for the quadrature angles of greatest violation of the inequalities. The value marked well $i$ refers to the product $\Pi V_{ij}$.}
\label{fig:EPRg1ki2}
\end{figure}

In Fig.~\ref{fig:EPRg1ki2} we show the EPR correlations for our second configuration, for $\chi=10^{-2}$. The results for $\chi=10^{-3}$ also show a violation of the inequalities, although not as great. In this configuration, we see that the quadrature angles for each well are closer to symmetric than in the first configuration examined here.

In conclusion, we have shown that experimental advances in the manipulation of potentials for condensed bosonic atoms have made possible another configuration which is a good candidate for the manufacture of continuous-variable non-Gaussian entangled states of massive particles. As our systems are both pumped and damped, they will reach a steady-state that will last as long as the pumping condensate is not significantly depleted. A system such as that which we analyse here is a further step toward bringing the flexibility of experimental quantum optics into the arena of quantum atom optics and can be readily expanded to more wells, with different pumping and damping configurations.

{\it Acknowledgments}

This research was supported by the Australian Research Council under the Future Fellowships Program (Grant ID: FT100100515).

\end{document}